\documentclass{iaus}
\usepackage{graphicx}
\usepackage{natbib}

\title
{Inner disk regions revealed by infrared interferometry}

\author
{Fabien Malbet}

\affiliation{Laboratoire d'Astrophysique de Grenoble, \break
  UJF/CNRS, BP 53, F-38041 Grenoble cedex 9, France \break
  Email: fabien.malbet@obs.ujf-grenoble.fr}  

\pubyear{2007}
\volume{243}  
\pagerange{}
\date{24/07/2007 and in revised form ??}
\jname{Star-disk interaction in young stars}
\editors{J. Bouvier and I. Appenzeller}

\newcommand{\microns}{\ensuremath{\mu\mbox{m}}}

\begin{document}

\maketitle

\begin{abstract}
  I review the results obtained by long-baseline interferometry at
  infrared wavelengths on the innermost regions around young stars.
  These observations directly probe the location of the dust and gas
  in the disks.  The characteristic sizes of these regions found are
  larger than previously thought. These results have motivated in part
  a new class of models of the inner disk structure. However the
  precise understanding of the origin of these low visibilities is
  still in debate. Mid-infrared observations have probed disk emission
  over a larger range of scales revealing mineralogy gradients in the
  disk. Recent spectrally resolved observations allow the dust and gas
  to be studied separately. The few results shows that the Brackett
  gamma emission can find its origin either in a wind or in a
  magnetosphere but there are no definitive answers yet. In a certain
  number of cases, the very high spatial resolution seems to reveal
  very close companions. In any case, these results provide crucial
  information on the structure and physical properties of disks
  surrounding young stars especially as initial conditions for planet
  formation.
\end{abstract}

\firstsection 
\section{Introduction}

Many physical phenomena occur in the inner regions of the disk which
surrounds young stars. The matter which falls onto the stellar surface
spirals in a more or less accreting circumstellar disk subject to
turbulence, convection, external and internal irradiation. The disks
which are rotating in a quasi Keplerian motion are probably the birth
location of future planetary systems.  Strong outflows, winds and even
jets find often their origin in the innermost regions of many young
stellar systems. The mechanisms of these ejection processes are not
well understood but they are probably connected to accretion. Most of young
stellar systems are born in multiple systems which can be very tight
and therefore have a strong impact on the physics of the disk inner
regions.

The details of all these physical processes are not yet well understood
because of lack of data to constrain them. The range
of physical parameters which define best the inner regions of disk in
young stellar objects are:
\begin{itemize}
\item radius ranging from 0.1\,AU to 10\,AU
\item temperature ranging from 150\,K to 4000\,K
\item velocities ranging from 10\,km/s to few 100\,km/s
\end{itemize}

The instrumental requirements to investigate the physical conditions
in such regions are therefore driven by the spectral coverage which
must encompass the near and mid infrared from 1 to $20\,\microns$.
Depending on the distance of the object (typically between 75\,pc and
450\,pc) the spatial resolution required to probe the inner parts of
disks ranges between fractions and a few tens of milli-arcseconds.
Since the angular resolution of astronomical instruments depends
linearly on the wavelength and inversely on the telescope diameter,
observing in the near and mid infrared wavelength domain points toward
telescopes of sizes ranging from ten to several hundreds of meters.
The only technique that allows such spatial resolution is therefore
infrared interferometry.

\citet{2007prpl.conf..539M}, published in \emph{Protostars and
  Planets}, review the main results obtained in infrared
interferometry in the domain of young stars between 1998 and 2005. The
purpose of the present review is to concentrate on the inner disk
regions and to give the latest results in this field. Section
\ref{sect:irinterf} briefly explains the principles of infrared
interferometry and lists the literature on the observations carried
out with this technique. Section \ref{sect:innerdisks} focuses on the
main results obtained on disk physics (sizes, structures, dust and gas
components,...) and Sect.~\ref{sect:others} presents results on other
phenomena constrained by interferometry (winds, magnetosphere,
multiple systems,...). In Sect.~\ref{sect:future} I finish the review
with the type of results that can be expected in the future.

\section{Infrared interferometry}
\label{sect:irinterf}

\subsection{Principle and observations}

Long baseline optical interferometry consists in mixing the light
received from an astronomical source and collected by several
independent telescopes separated from each other by tens or even
hundreds of meters. The light beams are then overlapped and form an
interference pattern if the optical path difference between the
different arms of the interferometer ---taking into account paths from
the source up to the detector--- is smaller than the coherence length
of the incident wave (typically of the order of several microns).
This interference figure is composed of fringes, i.e.\ a succession of
stripes of faint (destructive interferences) and bright (constructive
interferences) intensity. By measuring the contrast of these fringes,
i.e.\ the normalized flux difference between the darkest and brightest
regions, information about the morphology of the observed
astronomical source can be recovered.
\begin{figure}[t]
  \centering
  \parbox{0.4\hsize}{%
    \includegraphics[width=\hsize]{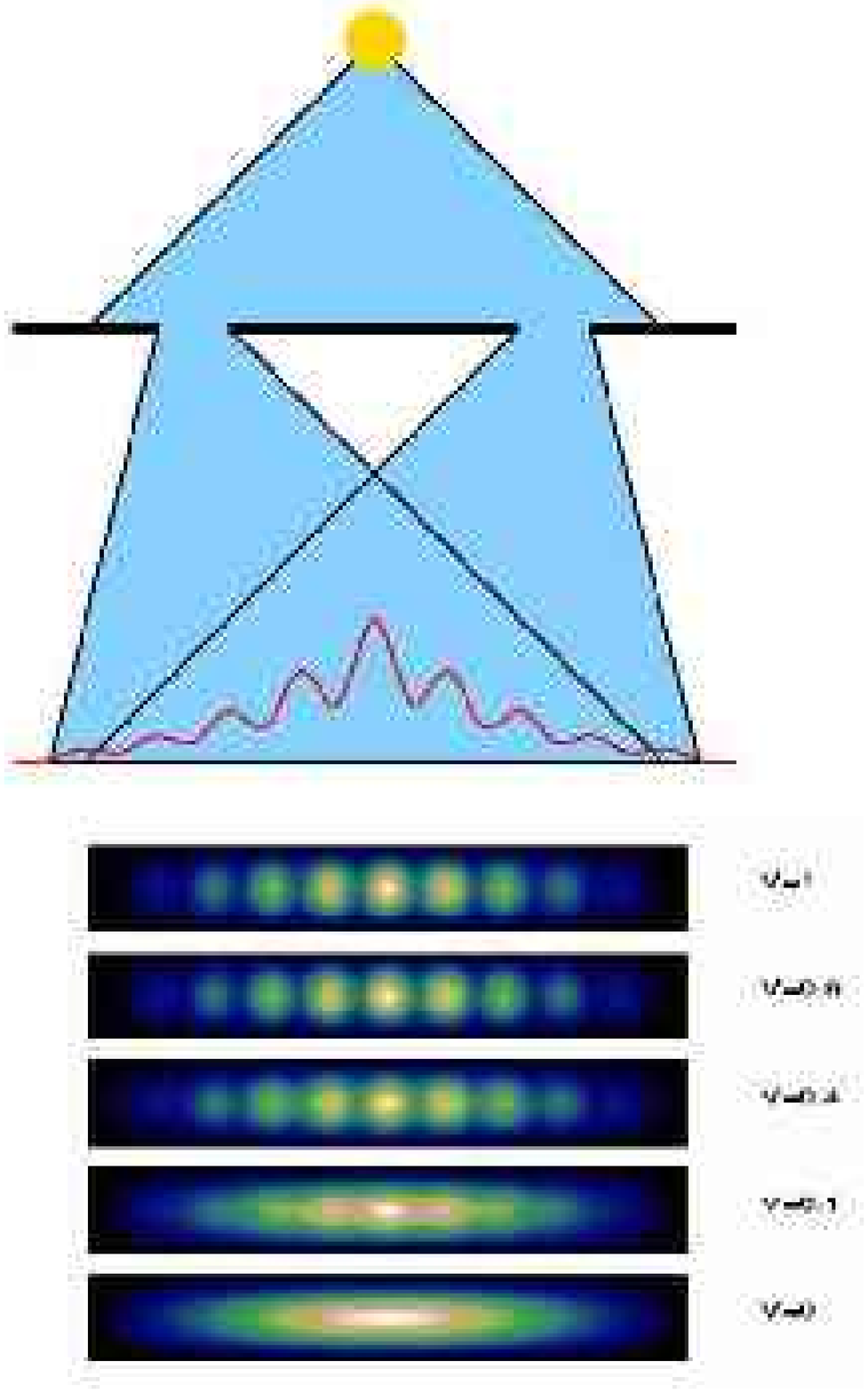}
  }%
  \parbox{0.4\hsize}{%
    \includegraphics[width=0.9\hsize]{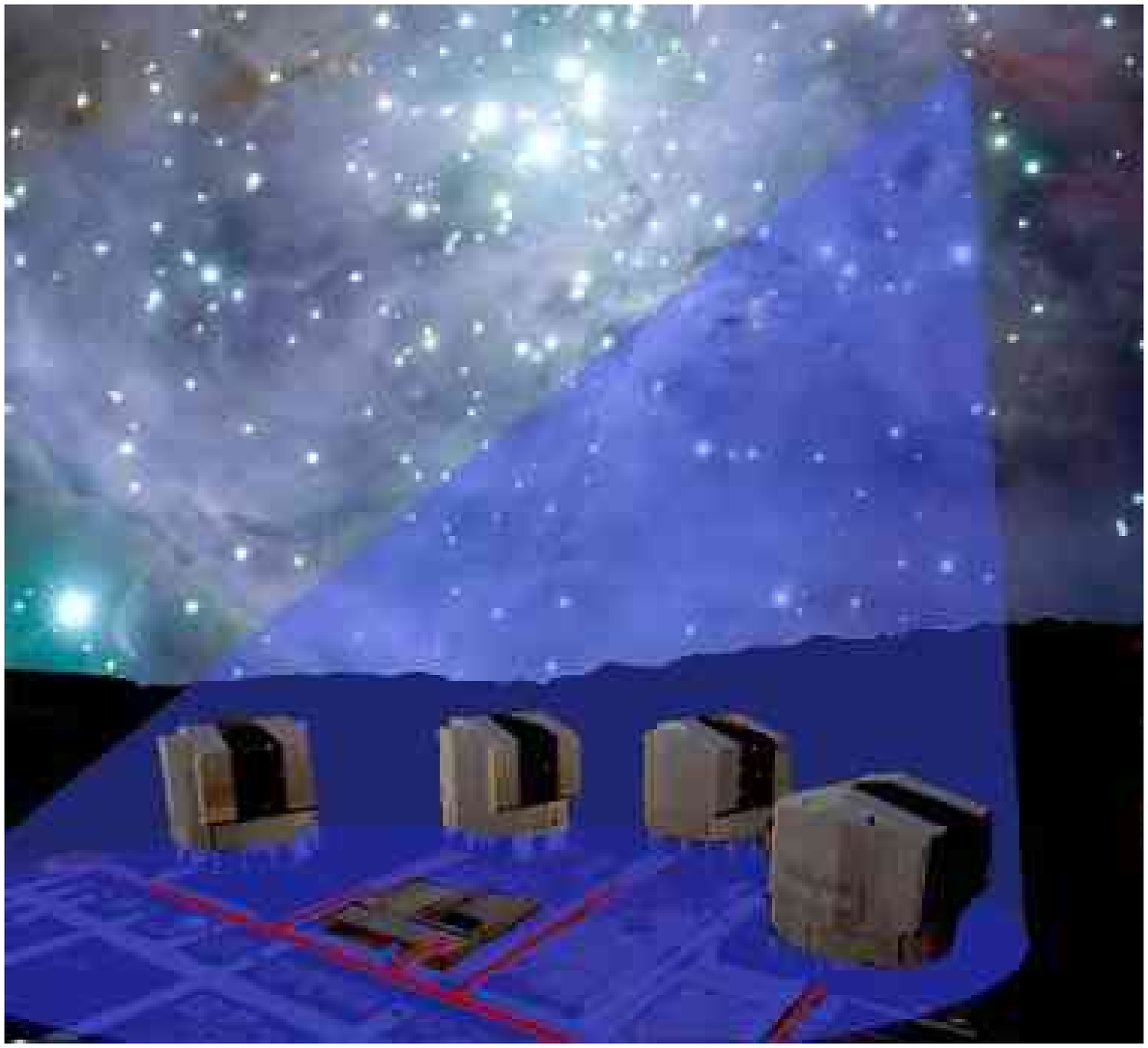}\\
    \includegraphics[width=0.9\hsize]{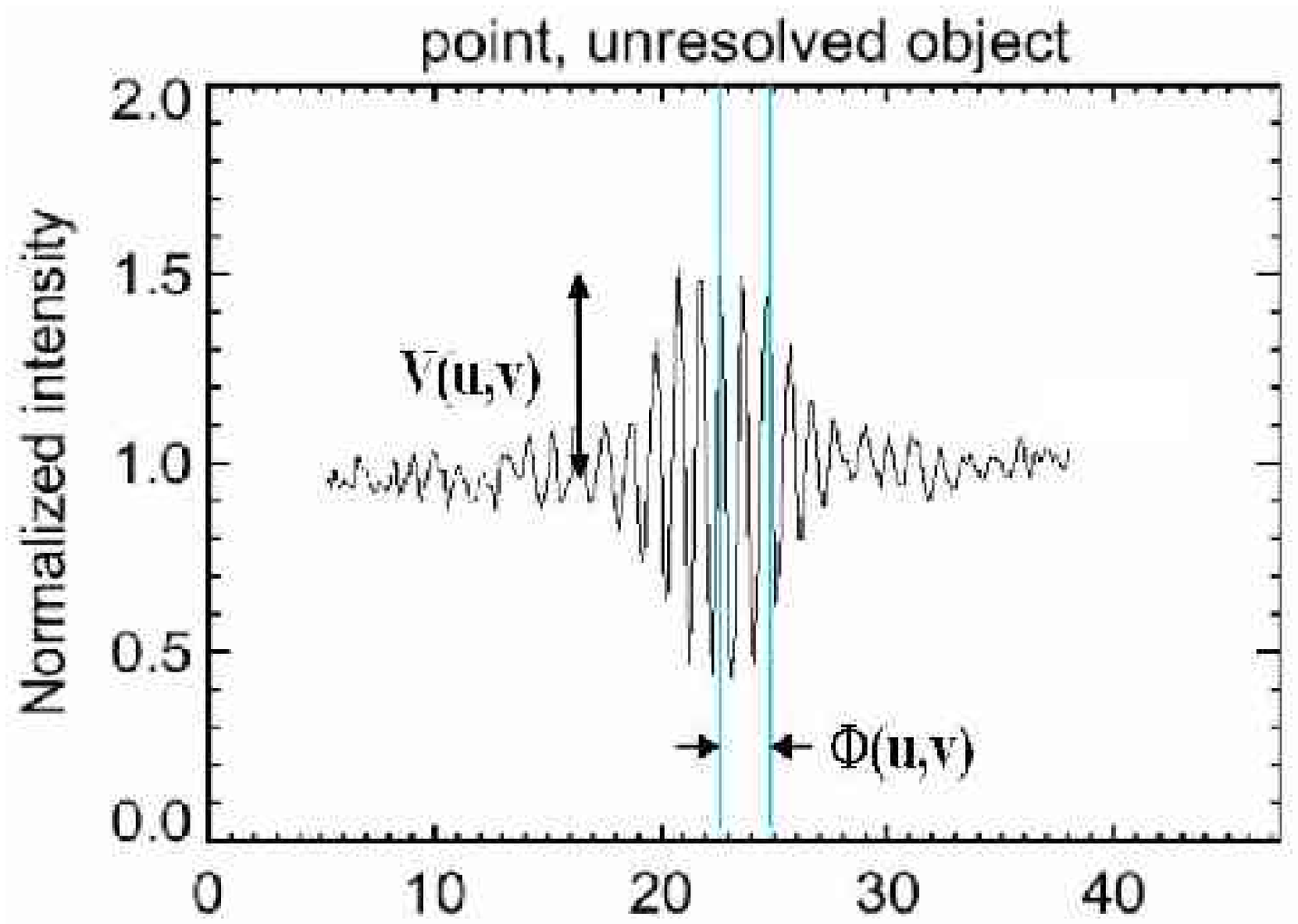}
  }\\
  \caption{Principle of interferometry. Upper panels: the Young's slit
    experiment (left) compared to optical interferometry (right): in
    both cases the light travels from a source to a plane where the
    incoming wavefront is split. The telescope apertures play the
    same role as the Young's slits. The difference lies in the propagation of
    light after the plane. In the case of optical
    interferometry, the instrument controls the propagation of light
    down to the detectors. At the detector plane, the light beams
    coming from the two apertures are overlapped. Lower panels:
    interference fringes whose contrast changes with the morphology of
    the source. Left panel shows fringes whose contrast varies from 0
    to 1. Right panel displays actual stellar fringes but scanned
    along the optical path. The measure of the complex visibilities
    corresponds to the amplitude of the fringes for the visibility
    amplitude and the position of the fringes in wavelength units for
    the visibility phase.}
  \label{fig:interf}
\end{figure}
Figure \ref{fig:interf} illustrates this principle.

\subsection{Instruments available for inner regions studies}

\begin{figure}[t]
  \centering
  \includegraphics[width=0.9\hsize]{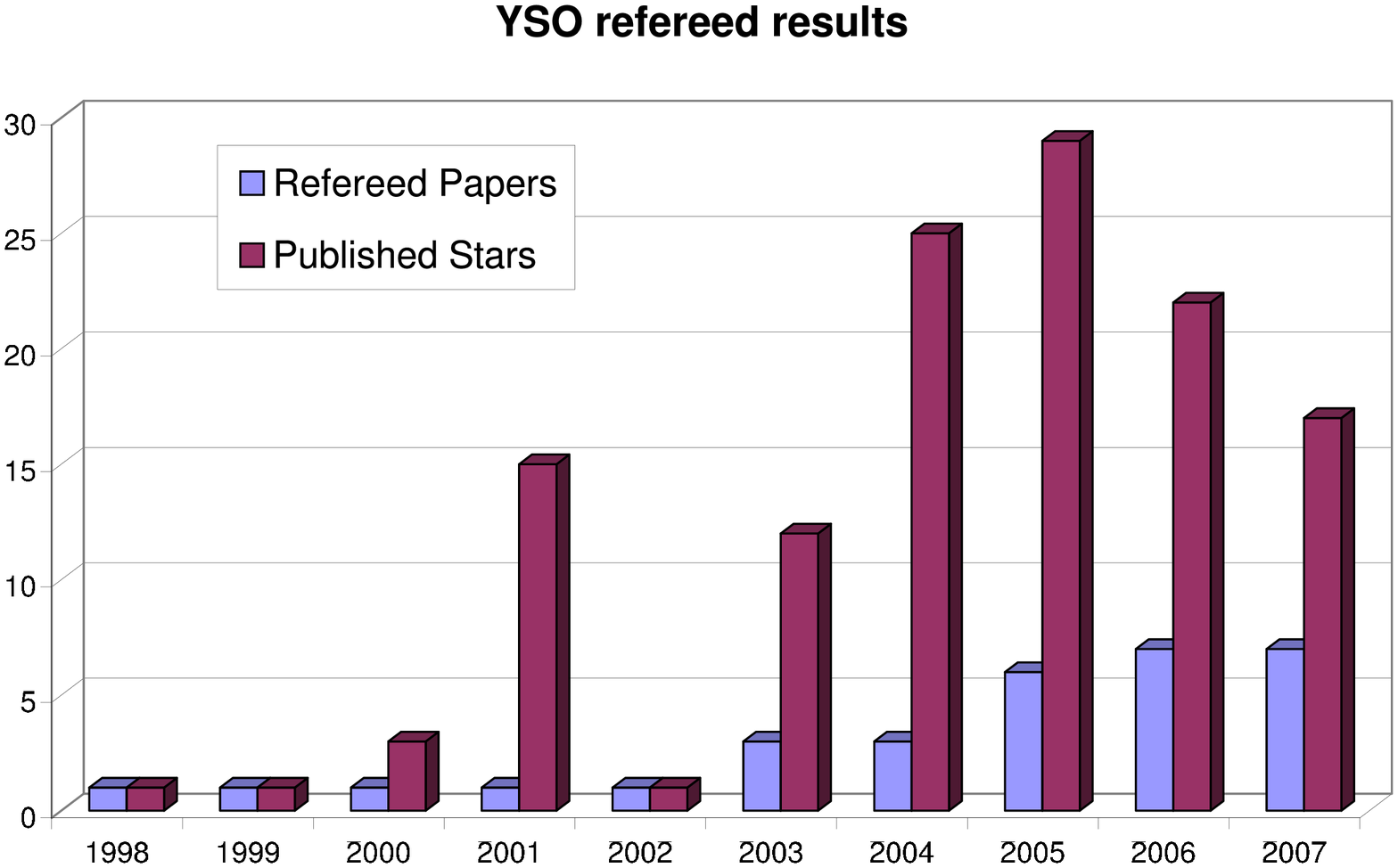}
  \caption{Young stellar objects observed by interferometry and number
    of refereed papers published in the period 1998-2007. The
    statistics of the year 2007 is not complete.}
  \label{fig:bib}
\end{figure}

\begin{figure}[t]
  \centering
  \includegraphics[width=0.9\hsize]{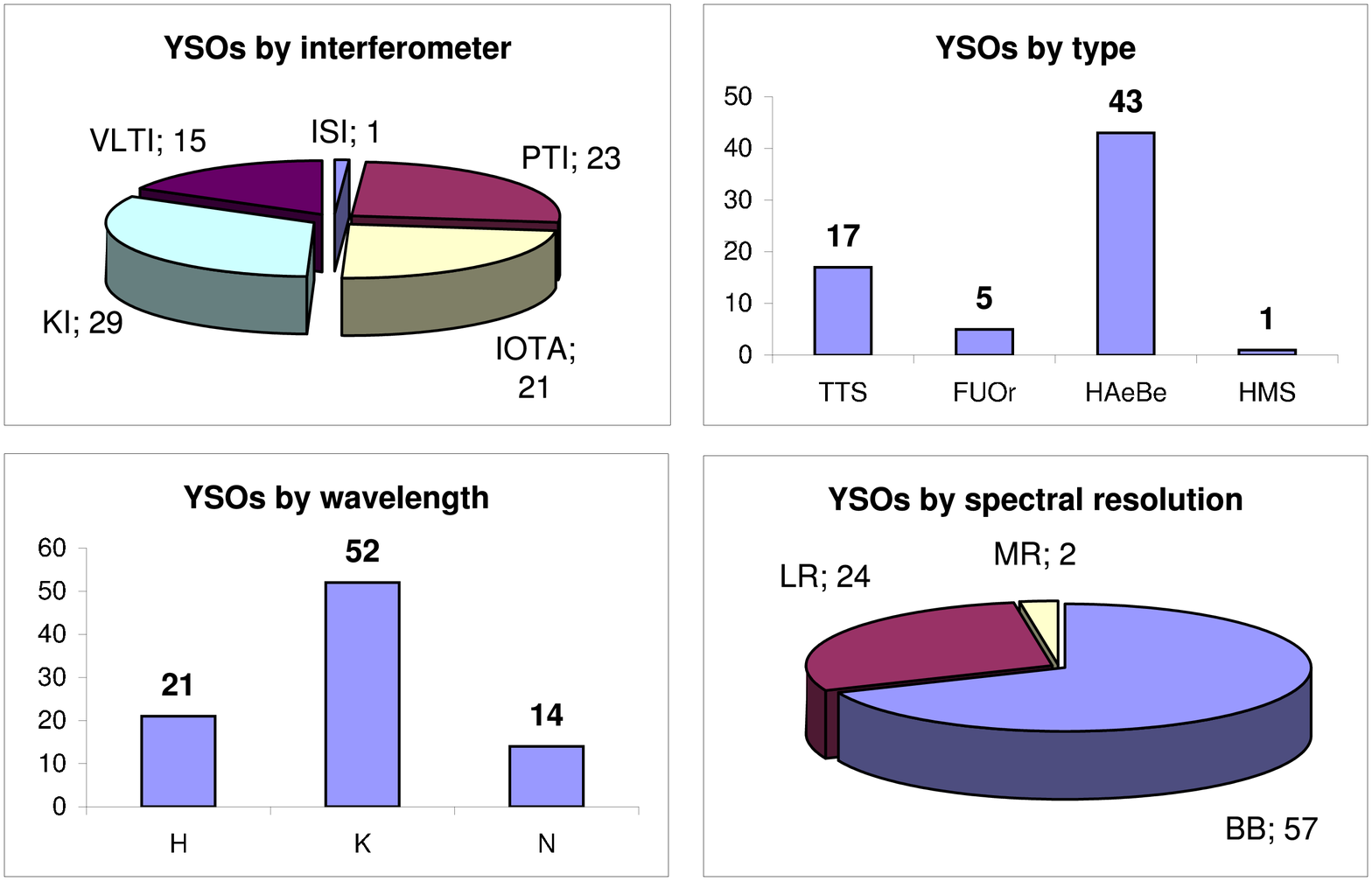}
  \caption{Young stellar objects observed by interferometry in the
    period 1998-2007. Upper left: distribution by
    interferometer. Upper right: distribution by YSO
    type. Lower left: distribution by wavelength of observation. Lower
  right: distribution by spectral resolution. The statistics is the
  same as the one of Fig.~\ref{fig:bib}. }
  \label{fig:various}
\end{figure}

\begin{table}[t]
  \renewcommand{\arraystretch}{1.25} 
  \centering
  \caption{Interferometers involved in YSO science}
  \label{tab:interferometers}
  \medskip
  \begin{tabular}{llllll}
    \hline\hline
    Facility &Instrument &Wavelength &Numbers of &Aperture    &Baseline\\
    &           &(microns)  &apertures  &diameter (m)&(m)\\
    \hline\hline
    PTI      &$V^2$      &$H$, $K$   &3          &0.4         &$80-110$\\
    \hline
    IOTA     &$V^2$, CP  &$H$, $K$   &3          &0.4         &$5-38$\\
    \hline
    ISI      &heterodyne &11         &2 (3)      &1.65        &$4-70$\\
    \hline
    KI       &$V^2$, nulling &$K$    &2          &10          &$80$\\
    \hline
    VLTI/AMBER &$V^2$, CP&$1-2.5$    &3 (8)      &8.2/1.8     &$40-130$\\
    &(imaging)  &/spectral   &           &            &/$8-200$\\
    \hline
    VLTI/MIDI&$V^2$ (/CP)&$8-13$     &2 (4)      &8.2/1.8     &$40-130$\\
    &           &/spectral  &           &            &/$8-200$\\
    \hline
    CHARA    &$V^2$, CP  &$1-2.5$    &2/4 (6)    &1           &$50-350$\\
    &(imaging)  &/spectral  &           &            &\\
    \hline
    LBT      &imaging,   &$1-10$     &2          &8.4         &$6-23$\\
    &nulling    &           &           &            &\\
    \hline\hline
    \multicolumn{6}{p{0.85\textwidth}}{\smallskip \footnotesize $V^2$:
      visibility measurement; CP: closure phase.}\\
    \multicolumn{6}{p{0.85\textwidth}}{\smallskip \footnotesize
      Acronyms. PTI: \emph{Palomar Testbed Interferometer}; 
      IOTA: \emph{Infrared and Optical Telescope Array} (closed since
      2006); ISI: \emph{Infrared Spatial Interferometer}; KI: \emph{Keck
        Interferometer}; VLTI: \emph{Very Large Telescope
        Interferometer}; CHARA: \emph{Center for High Angular
        Resolution Array}; LBT: \emph{Large Binocular Telescope} (not
      yet operational).}\\
  \end{tabular}
\end{table}
Interferometric observations of young stellar objects were and are
still performed at six different facilities on seven different
instruments (see Table~\ref{tab:interferometers}). We can classify
these observations into three different categories:
\begin{itemize}
\item \textbf{Small-aperture interferometers}: PTI, IOTA and ISI were
  the first facilities to be operational for YSO observations in the
  late 1990's (see Figs.~\ref{fig:bib} and \ref{fig:various}). They
  have provided mainly the capability of measuring visibility
  amplitudes and lately closure phases. The latest one, CHARA, has
  aperture diameter of 1\,m. The instruments are mainly
  accessible through team collaboration.
\item \textbf{Large-aperture interferometers}: KI, VLTI and soon LBT
  are facilities with apertures larger than 8\,m. The instruments are
  widely open to the astronomical community through general calls for
  proposals. Lately, these facilities have significantly increased the
  number of young objects observed.
\item \textbf{Instruments with spectral resolution:} CHARA, MIDI and
  AMBER provides spectral resolution from a few hundred up to 10,000
  whereas other instruments mainly provided broadband observations.
  The spectral resolution allows the various phenomena occurring in the
  environment of young stars to be separated.
\end{itemize}


\subsection{Elements of bibliography}

Figure \ref{fig:bib} displays the number of published results, and
show that it is increasing with time and improved facilities. At the
date of the conference there were 31 refereed articles published in
the field of young stars corresponding to 66 young stellar objects
observed \citep[see by chronological order:][]{1998ApJ...507L.149M,
  1999ApJ...513L.131M, 2000ApJ...543..313A, 2001ApJ...546..358M,
  2002ApJ...577..826T, 2003ApJ...588..360E, 2003ApJ...592L..83C,
  2003Ap&SS.286..145W, 2004A&A...423..537L, 2004ApJ...613.1049E,
  2004Natur.432..479V, 2005A&A...437..627M, 2005ApJ...622..440A,
  2005ApJ...623..952E, 2005ApJ...624..832M, 2005ApJ...635..442B,
  2005ApJ...635.1173A, 2006ApJ...637L.133E, 2006ApJ...641..547M,
  2006A&A...458..235P, 2006A&A...449L..13A, 2006ApJ...645L..77M,
  2006ApJ...647..444M, 2006ApJ...648..472Q, 2007ApJ...657..347E,
  2007A&A...464...43M, 2007A&A...464...55T, 2007A&A...466..649K,
  2007A&A...469..587L, 2007Natur.447..562E, 2007A&A...471..173R}.

Graphs in Fig.~\ref{fig:various} show that the distribution of
observed object is rather well distributed among the various
facilities. Several categories of young stellar systems have been
observed at milli-arcsecond scales mainly in the near-infrared
wavelength domain, but also in the mid-infrared one. They include the
brightest Herbig Ae/Be stars, the fainter T Tauri stars and the few FU
Orionis. Finally most observations were carried out in broad band but
the advent of large aperture interferometers like the VLTI and KI
allow higher spectral resolution to be obtained.

\section{Inner disk physics}
\label{sect:innerdisks}

Most of the studies carried out on YSOs are focused on the physics of
inner regions of disks. They started with the determination of rough sizes
of emission then led to more constraints on the disk structure. Mid
infrared spectrally resolved observations were able to identify
different types of dust grains. Near infrared spectrally resolved
observations are coming out and permits to spatially discriminate
between gas and dust.  

\subsection{Sizes of circumstellar structures}

About 10 years ago, the paradigm was that disks were present around a
majority of young stars. These disks were believed to behave
``normally'' with a radial temperature distribution following a
power-law $T \propto r^{-q}$ with q ranging between 0.5 and 0.75. The
value of $q$ depends on the relative effect of irradiation from the
central star in comparison with heat dissipation due to accretion.
This model was successful to reproduce ultraviolet and infrared
excesses in spectral energy distributions (SEDs).
\citet{1995A&AS..113..369M} investigated the potential of optical long
baseline interferometry to study the disks of T Tauri stars and FU
Orionis stars. They found that the structure would be marginally
resolved but observations would be possible with baselines of the
order of 100\,m with a visibility amplitude remaining high.

First observations of brighter Herbig Ae/Be stars showed that the
observed visibilities were much smaller than the expected ones
especially in these objects were the accretion plays a little role.
\citet{2002ApJ...579..694M} pointed out that the interferometric sizes
of these objects were much larger than expected in the standard disk
model. They plotted the sizes obtained in function of the stellar
luminosity and found out that there was a strong correlation following
a law in $L^{0.5}$ over two decades. This behavior is consistent with
the variation of radius of dust sublimation with respect to the central
star luminosity: the more luminous the object the further the dust can
survive in solid form because of temperature lower than the
sublimation limit ($\sim 1000-1500\,\mbox{K}$). Only the most massive
Herbig Be stars seem to be compliant with the standard accretion disk
model.

In the meantime, in order to account for the near-infrared
characteristics of SEDs and in particular a flux excess around
$\lambda=3\microns$, \citet{2001A&A...371..186N} proposed that disks
around Herbig Ae/Be stars have optically thin inner cavity and
create a puffed-up inner wall of optically thick dust at the dust
sublimation radius. More realistic models were developed afterward
which takes more physical properties into account
\citep{2001ApJ...560..957D, 2004ApJ...617..406M,
  2005A&A...438..899I}. However as pointed out by
\citet{2003MNRAS.346.1151V} the models are not the unique ones that can
reproduce the measurements and they proposed another model with a disk
halo.

Observations at KI \citep{2003ApJ...592L..83C, 2005ApJ...623..952E,
  2005ApJ...622..440A} found also large NIR sizes for lower-luminosity
T Tauri stars, in many cases even larger than would be expected from
extrapolation of the HAe relation. It is interpreted by the fact that
the accretion disk contributes significantly to the luminosity emitted
by the central region and therefore this additional luminosity must be
taken into account in the relationship . However in these systems the
error bars are still large and very few measurements have been obtained per
object. In order to interpret all the T Tauri measurements,
\citet{2005ApJ...622..440A} need to introduce new physical phenomena
like optical thick gas emission in the inner hole and extended
structure around the objects.

Characteristic dimensions of the emitting regions at $10\,\microns$
were found by \cite{2004A&A...423..537L} to be ranging from 1 AU to 10
AU. The sizes of their sample stars correlated with the slope of the
$10-25\,\microns$ infrared spectrum: the reddest objects are the
largest ones.  Such a correlation is consistent with a different
geometry in terms of flaring or flat (self-shadowed) disks for sources
with strong or moderate mid-infrared excess, respectively,
demonstrating the power of interferometry not only to probe
characteristics sizes of disk but also to derive information on the
vertical disk structure.

\subsection{Constraints on disk structure}
\label{sect:diskstructure}

Theoreticians start discussing slightly different scenarios of the
inner regions around young stars. For example, the shape of the inner
puffed-up wall is modeled with a curved shape by
\citet{2005A&A...438..899I} due to the very large vertical density
gradient and the dependence of grain evaporation temperature on gas
density as expected when a constant evaporation temperature is
assumed. Recently, \citet{2007ApJ...661..374T} proposed that the
geometry of the rim depends on the composition and spatial
distribution of dust due to grain growth and settling.

\citet{2007ApJ...658..462V} presented a model-independent method of
comparison of NIR visibility data of YSOs. The method based on scaling
the measured baseline with the YSO distance and luminosity removes the
dependence of visibility on these two variables. They found that low
luminosity Herbig Ae stars are best explained by the uniform
brightness ring and the halo model, T Tauri stars with the halo model,
and high luminosity Herbig Be stars with the accretion disk model, but
they admit that the validity of each model is not well established.

At the current moment, only one object has been thoroughly studied: FU Orionis
\citep{1998ApJ...507L.149M, 2005A&A...437..627M, 2006ApJ...648..472Q}.
This young stellar object has been observed on 42 nights over a period
of 6 years from 1998 to 2003 with 287 independent measurements of the
fringe visibility at 6 different baselines ranging from 20 to 110 m in
length, in the $H$ and $K$ bands. The data not only resolves FU Ori at
the AU scale, but also allows the accretion disk scenario to be
tested. The most probable interpretation is that FU Ori hosts an
active accretion disk whose temperature law is consistent with the
standard model. In the mid infrared, \citet{2006ApJ...648..472Q}
resolved structures that are also best explained with an optically
thick accretion disk. A simple accretion disk model fits the observed
SED and visibilities reasonably well and does not require the presence
of any additional structure such as a dusty envelope. This is why one
should remain careful with results coming from surveys having only few
measurements per object. 

\citet{2006ApJ...641..547M} obtained $K$-band observations of three
other FU Orionis objects, V1057 Cyg, V151 Cyg, and Z CMa-SE and found
that all three objects appear significantly more resolved than
expected from simple models of accretion disks tuned to fit the SEDs.
They believe that emission at the scale of tens of AU in the
interferometer field of view is responsible for the low visibilities,
originating in scattering by large envelopes surrounding these
objects. In a not yet published study, Li Causi et al.\ have measured
again interferometric visibilities of Z CMa with VLTI/AMBER and
propose to interpret the data by the presence of a very close
companion.

\begin{figure}[t]
  \centering 
  \includegraphics[width=\hsize]{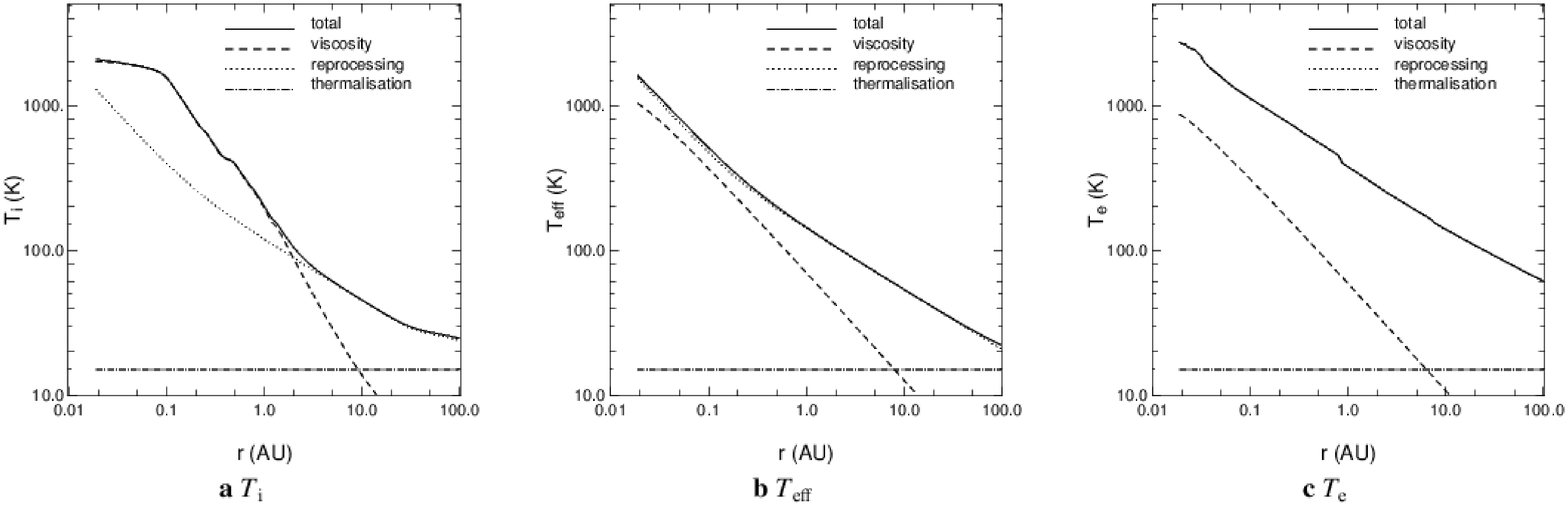}
  \caption{Radial distribution of temperature at different location in
    the disk computed with a two-layer model by
    \citet{2003A&A...400..185L}. Left: temperature in the equatorial
    plane. Center: effective temperature. Right: surface
    temperature. The different lines represent the various
    contribution to the heating: viscosity, reprocessing of the
    stellar light, thermalization with the ambient temperature.}
  \label{fig:lmm2003}
\end{figure}

On the pure theoretical side, very few physical models achieved to
fit interfermoetric data simultaneously with SEDs. Using a two-layer
accretion disk model, \citet{2003A&A...400..185L} found satisfactory
fits for SU Aur, in solutions that are characterized by the midplane
temperature being dominated by accretion, while the emerging flux is
dominated by reprocessed stellar photons (see Fig.~\ref{fig:lmm2003}).
Since the midplane temperature drives the vertical structure of the
disk, there is a direct impact on the measured visibilities that are
not necessarily taken into account by other models.

Very interesting results have been presented in this conference by
Kraus et al. (this volume) showing that they are able to derive the
temperature radial distribution of the disk around MWC 147 from the
interferometric measurements using the spectral variation of the
visibilities in low resolution. A similar work has been attempted at
PTI with larger error bars \citep{2007ApJ...657..347E}.   

\subsection{Dust mineralogy}

The mid-infrared wavelength region contains strong resonances of
abundant dust species, both oxygen-rich (amorphous or crystalline
silicates) and carbon-rich (polycyclic aromatic hydrocarbons,
or PAHs). Therefore, spectroscopy of optically thick
protoplanetary disks offers a diagnostic of the chemical
composition and grain size of dust in disk atmosphere. 

\citet{2004Natur.432..479V} spatially resolved three protoplanetary
disks surrounding Herbig Ae/Be stars across the $N$ band. The
correlated spectra measured by MIDI at the VLTI correspond to disk
regions ranging from 1 to 2\,AUs. By combining these measurements with
unresolved spectra, the spectrum corresponding to outer disk regions
at 220 AU can also be derived. These observations have revealed that
the dust in these regions was highly crystallized (40 to 100\%), more
than any other dust observed in young stars until now. The spectral
shape of the inner-disk spectra shows surprising similarity with Solar
System comets. Their observations imply that silicates crystallize
before terrestrial planets are formed, consistent with the
composition of meteorites in the Solar System. Similar measurements
were also carried out by \citet{2007A&A...471..173R} on the T Tauri
system, TW Hya.  According to the correlated flux measured with MIDI,
most of the crystalline material is located in the inner, unresolved part of
the disk, about 1\,AU in radius.

\subsection{Gas/dust connection}

\citet{2005astro.ph..8052G} observed the young stellar system 51 Oph
confirming the interpretation of \citet{2005A&A...430L..61T} and
more recently \citet{2007ApJ...660..461B} of a disk seen edge-on: 
the radial distribution of excitation temperatures for the
vibrational levels of CO overtone ($\Delta v=2$) emission from hot gas
is consistent with the gas being in radiative thermal equilibrium
except at the inner edge, where low vibrational bands have higher
excitation temperatures. In not yet published results,
Tatulli et al. (priv.\ comm.) confirm the high inclination of the disk
but also detect the CO bandheads allowing the dust responsible for the
continuum to be separated from the gas emitting this CO bands. As a
matter of fact, the visibilities in the CO bands is lower than the
ones measured in the continuum implying that the region responsible
for this gas emission is smaller than the region responsible for the
dust emission. 
\begin{figure}
  \centering
  \includegraphics[width=0.5\hsize]{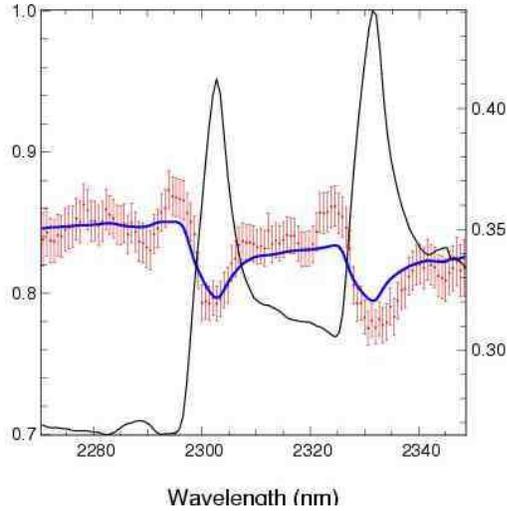}
  \caption{Spectrally dispersed visibility amplitudes of 51 OPh in the
    CO bandhead spectral region. Overimposed is the spectrum as
    measured by VLTI/AMBER (black line). The blue curve corresponds to
    the addition of simple uniform disk model for the excess emission
    in the line with a typical dimaeter of 0.2\,AU. From Tatulli et
    al.\ (priv.\ comm.)}
  \label{fig:51oph}
\end{figure}
Figure \ref{fig:51oph} illustrates this result.  

This result shows that the combination of very high spatial information
with spectral resolution opens brand new perspectives in the studies of the
inner disk properties by discriminating between species. 

\section{Other AU-scale phenomena}
\label{sect:others}

Several other physical phenomena have been investigated in the
innermost region of disks: wind, magnetosphere and close companions.

\subsection{Outflows and winds}

The power of spectrally resolved interferometric measurements provides
detailed wavelength dependence of inner disk continuum emission (see
end of Sect.~\ref{sect:diskstructure}). These new capabilities enable
also detailed studies of hot winds and outflows, and therefore the
physical conditions and kinematics of the gaseous components in which
emission and absorption lines arise like Br$\gamma$ and H$_2$ lines.
With VLTI/AMBER, \citet{2007A&A...464...43M} spatially
resolved the luminous Herbig Be object MWC 297, measuring visibility
amplitudes as a function of wavelength at intermediate spectral
resolution R = 1500 across a $2.0-2.2\,\microns$ band, and in
particular the Br$\gamma$ emission line. The interferometer
visibilities in the Br$\gamma$ line are about 30\% lower than those of
the nearby continuum, showing that the Br$\gamma$ emitting region is
significantly larger than the NIR continuum region.  Known to be an
outflow source, a preliminary model has been constructed in which a
gas envelope, responsible for the Br$\gamma$ emission, surrounds an
optically thick circumstellar disk.  The characteristic size of the
line-emitting region being 40\% larger than that of the NIR disk.
This model is successful at reproducing the VLTI/AMBER measurements as
well as previous continuum interferometric measurements at shorter and
longer baselines \citep{2001ApJ...546..358M, 2004ApJ...613.1049E}, the
SED, and the shapes of the H$\alpha$, H$\beta$, and Br$\gamma$
emission lines.  The precise nature of the MWC 297 wind, however,
remains unclear; the limited amount of data obtained in these first
observations cannot, for example, discriminate between a stellar or
disk origin for the wind, or between competing models of disk winds
(e.g. Ferreira et al.\ 2007 and Shu et al. 2007 both in this volume).

\subsection{Magnetosphere}

The origin of the hydrogen line emission in Herbig Ae/Be stars is
still unclear. The lines may originate either in the gas which
accretes onto the star from the disk, as in magnetospheric accretion
models \citep{1994ApJ...426..669H}, or in winds and jets, driven by
the interaction of the accreting disk with a stellar
\citep{1994ApJ...429..781S} or disk \citep{2000A&A...353.1115C}
magnetic field. For all models, emission in the hydrogen lines is
predicted to occur over very small spatial scales, a few AUs at most.
To understand the physical processes that happen at these scales, one
needs to combine very high spatial resolution with enough spectral
resolution to resolve the line profile. 

One one hand, \citet{2007A&A...464...55T} performed interferometric
observations of the Herbig Ae star HD~104237, obtained with the
VLTI/AMBER instrument with $R = 1500$ high spectral resolution. The
observed visibility was identical in the Br$\gamma$ line and in the
continuum, even though the line represents 35\% of the continuum flux.
This immediately implies that the line and continuum emission regions
have the same apparent size.  Using simple toy models to describe the
Br$\gamma$ emission, they showed that the line emission is unlikely to
originate in either magnetospheric accreting columns of gas or in the
gaseous disk but more likely in a compact outflowing disk wind
launched in the vicinity of the rim, about 0.5\,AU from the star. The
main part of the Br$\gamma$ emission in HD~104237 is unlikely to
originate in magnetospheric accreting matter.

On the other hand, \citet{2007Natur.447..562E} measured an increase of
the Br$\gamma$ visibility in MWC 480 implying that the region of
emission of the hydrogen line is very compact, less than 0.1\,mas in
radius which could be interpreted as an emission originated in the
magnetosphere of the system.  

At the present time, given the limited number of samples, it is
difficult to derive a general tendency but it seems that all possible
scenari can be found.

\subsection{Binaries and multiple systems}

\citet{2005ApJ...635..442B} performed the first direct measurement
of pre-main sequence stellar masses using interferometry,
for the double-lined system HD 98800-B. These authors established
a preliminary orbit that allowed determination of the
(subsolar) masses of the individual components with 8\% accuracy.
Comparison with stellar models indicates the need for
subsolar abundances for both components, although stringent
tests of competing models will only become possible
when more observations improve the orbital phase coverage
and thus the accuracy of the stellar masses derived.

Another example, based on a low-level oscillation in the visibility
amplitude signature in the PTI data of FU Ori,
\citet{2005A&A...437..627M} claim the detection of an off-centered
spot embedded in the disk that could be physically interpreted as a
young stellar or protoplanetary companion located at $\sim 10\,\mbox{AUs}$, and could
possibly be at the origin of the FU Ori outburst itself. Using another
technique, \citet{2006ApJ...645L..77M} reported on the detection of
localized off-center emission at 1-4 AU in the circumstellar
environment of AB Aurigae. They used closure-phase measurements in the
near-infrared. When probing sub-AU scales, all closure phases are
close to zero degrees, as expected given the previously determined
size of the AB Aurigae inner-dust disk. However, a clear closure-phase
signal of $-3.5^{\circ}\pm0.5^{\circ}$ is detected on one triangle
containing relatively short baselines, requiring a high degree of
asymmetry from emission at larger AU scales in the disk. They
interpret such detected asymmetric near-infrared emission as a result
of localized viscous heating due to a gravitational instability in the
AB Aurigae disk, or to the presence of a close stellar companion or
accreting substellar object.

\section{Future prospects and conclusion}
\label{sect:future}

As emphasized in this review, more interferometric data is required
with better accuracy and also wider coverage of the baselines in order
to constrain better the models that have been proposed. Like for radio
astronomy, these supplementary data will allow image reconstruction
without any prior knowledge of the observed structure. Several
projects are already ready to obtain interferometric images although
with few pixels across the field: MIRC at CHARA and AMBER at the VLTI
in the near-infrared. However at the moment MIRC is limited in
sensitivity and AMBER in number of telescopes (3) which makes it
difficult to routinely achieve imaging. In the mid-infrared the
MATISSE instrument is being proposed to ESO to provide imaging with 4
telecopes at the VLTI. VSI is also a proposed VLTI instrument of
second generation which can combine from 4 to 8 beams at the same time
so that imaging becomes easier. LBT will also provides imaging capability.

All these instruments provide spectral resolution that make them
indeed spectro-imager. Therefore in the future, one should be able to
obtain a wealth of information from the innermost regions of disks
around young stars. However in the meantime, observations are already
mature enough to allow detailed modeling of the phenomena occuring in
these inner regions.

\bibliography{malbet-iaus243}
\bibliographystyle{aa}

\end{document}